\documentclass[aps,pra,twocolumn,showpacs,groupedaddress]{revtex4}
\usepackage{amsfonts}
\usepackage{amsmath}
\usepackage{amssymb}
\usepackage{graphicx}
\newcommand{\Tr}{\text{Tr}}
\newcommand{\ket}[1]{|#1\rangle}
\newcommand{\bra}[1]{\langle#1|}

\newcommand{\schrodinger}{Schr\"{o}dinger }

\newcommand{\sigmam}{\sigma^-}
\newcommand{\sigmap}{\sigma^+}
\newcommand{\inner}[2]{\langle #1|#2\rangle}
\newcommand{\ext}[2]{|#1\rangle\langle#2|}

\newcommand{\pt}[1]{\frac{\partial}{\partial t}#1}
\newcommand{\dt}[1]{\frac{d}{d t}#1}

\newcommand{\PRA}[3] {Phys. Rev. A {\bf #1}, #2
(#3)}
\newcommand{\PRE}[3] {Phys. Rev. E {\bf #1}, #2
(#3)}

\newcommand{\JPB}[3] {J. Phys. B {\bf #1}, #2 (#3)}
\newcommand{\PLA}[3] {Phys. Lett. A {\bf #1}, #2 (#3)}
\newcommand{\EPL}[3] {Europhysics Letters {\bf #1}, #2 (#3)}
\newcommand{\JMO}[3] {J. Mod. Opt. {\bf #1}, #2 (#3)}
\newcommand{\PS}[3] {Phys. Scr. {\bf #1}, #2 (#3)}

\begin{document}

\title{Effective Hamiltonian Approach to Open Systems and Its Applications}
\author{X. L. Huang$^1$, X. X. Yi$^{1,2}$\footnote{yixx@dlut.edu.cn},
Chunfeng Wu$^2$, X. L. Feng$^2$, S. X. Yu$^2$, and C. H.
Oh$^2$\footnote{phyohch@nus.edu.sg}} \affiliation{$^1$School of
Physics and Optoelectronic Technology, Dalian University of
Technology, Dalian 116024 China } \affiliation{$^2$Centre for
Quantum Technologies and Department of Physics, National University
of Singapore, 3 Science Drive 2, Singapore 117543, Singapore}
\date{\today}

\begin{abstract}
By using the effective Hamiltonian approach, we present a
self-consistent framework for the analysis of geometric phases and
dynamically stable decoherence-free subspaces in open systems.
Comparisons to the earlier works are made. This effective
Hamiltonian approach is then extended  to a non-Markovian case with
the generalized Lindblad master equation. Based on this extended
effective Hamiltonian approach, the non-Markovian  master equation
describing a dissipative two-level system is solved, an adiabatic
evolution is defined and the corresponding adiabatic condition is
given.
\end{abstract}

\pacs{ 03.65.Bz, 03.65.Ta, 07.60.Ly} \maketitle

\section{Introduction\label{sectionintroduction}}
 The evolution of a system that interacts with its surrounding
 environment is fully given by a dynamical map that corresponds to a quantum
stochastic process. The state, the environment and their
correlations change with time. If the environment is assumed not to
react on the system with memory, the Markov approximation can be
taken in which these correlations are discarded to derive the
Kossakowski-Lindblad master
equation\cite{kossakowski72,gorini76,Lindblad}. This theory extends
quantum mechanics beyond Hamiltonian dynamics,  and find powerful
applications in quantum optics\cite{quantumoptics} and quantum
information\cite{Nielsen}. Many
approaches\cite{solution1,solution2,Krausmethod,
Liemethod1,Liemethod2,perturbativeYi,effectivesolution,dampingbasis}
have been proposed  to solve this Lindblad master equation,
including  the Kraus representation\cite{Krausmethod},
 the Lie algebra approach\cite{Liemethod1,Liemethod2}, the
perturbative expansion\cite{perturbativeYi}, the approach based on
the damping bases\cite{dampingbasis} and the effective Hamiltonian
approach\cite{effectivesolution}. The key idea of the effective
Hamiltonian approach is to map the Lindblad master equation to a
\schrodinger equation by introducing an ancilla, this leads to the
advantage that almost all methods developed to solve(or analyze) the
\schrodinger equation can be borrowed to solve(or analyze)  the
master equation. Leveraging on  this advantage to define and
calculate the geometric phase, as well as to formulate the
dynamically decoherence-free subspaces\cite{karasik08} is one goal
of this paper.

The Markov approximation is inadequate  for many physical phenomena.
Even if the environment is large compared to the system, it might
still react on the system with memory. For example, the system can
only couple to a few environmental degrees of freedom for short
times, resulting in a memory effect of the environment on the
system. In fact, short time scales in experiments often show
environmental memory effects, a decay that can be partially undone
by exploiting environmental memory effects\cite{hahn50} in the case
of spin-echoes is an example. Also, non-Markovian quantum effects
may play a role in the energy transfer in
photosynthesis\cite{engel07}. Hence modeling non-Markovian open
quantum systems is crucial for understanding these experiments.
There are many extensions developed to  go beyond the Markov
approximation\cite{shabani05,uchiyama06,
breuer07,expmemoryhazard,expmemorypositivity,
postMarkovian,postMarkoviansolution1,postMarkoviansolution2}. Among
them, Breuer\cite{breuer07} has derived  a non-Markovian master
equation by using  the correlated projection superoperators
technique\cite{BPbookprojection,projection,nonMarkovianPRE,ferraro08,vacchini08},
this master equation can be written in a generalized Lindblad form
and thus it is local in time. The other goal of this paper is to
extend the effective Hamiltonian approach to  this generalized
Lindblad master equation,  an  extended damping basis for this
non-Markovian dynamics  is also given. A connection between these
two approaches is presented.

This paper is organized as follows. In Sec.{\rm II}, after
introducing the effective Hamiltonian approach, we present a
definition for the adiabatic and non-adiabatic geometric phase for
the open system, a connection  to the earlier work is established.
Based on the effective Hamiltonian approach, we formulate the
dynamically stable decoherence-free subspaces, necessary and
sufficient conditions for the dynamically stable decoherence-free
subspace are provided and discussed. This effective Hamiltonian
approach is extended  to a non-Markovian case in Sec.{\rm III}. An
example to calculate the effective Hamiltonian and solve the
non-Markovian master equation is given in Sec.{\rm IV}. We apply
this generalized effective Hamiltonian approach to analyze the
decoherence-free subspaces and define the adiabatic evolution for
this non-Markovian dynamics in Sec.{\rm V}. Finally, we conclude our
results in Sec.{VI}.
\section{Geometric phase and decoherence-free subspaces in the
effective Hamiltonian approach}
 The effective Hamiltonian approach\cite{effectivesolution,effectiveadiabatic}
is a  method to solve the Lindblad master equation. The main idea of
this method can be  outlined as follows. By introducing an ancilla,
which has the same dimension of Hilbert space  as the system, we can
map the system density matrix $\rho(t)$ to a wave function of the
composite system (system + ancilla). A \schrodinger-like equation
can be derived from the master equation. The solution of the master
equation can be obtained by mapping the solution of the
\schrodinger-like equation back to the density matrix. Assume the
dimension of the Hilbert space for both the system and the ancilla
is $N$, and let $\ket{E_n(0)}$ and $\ket{e_m(0)}$ denote the
eigenstates for the system  and the ancilla, respectively. The
mathematical representation of the above idea can be formulated  as
follows.  A wave function for the composite system in the
$N^2$-dimensional Hilbert space may be constructed as
\begin{eqnarray}
\rho(t)\rightarrow\ket{\Psi(t)}=
\sum_{m,n=1}^N\rho_{mn}(t)\ket{E_m(0)}\ket{e_n(0)},\label{map}
\end{eqnarray}
where $\rho_{mn}(t)=\bra{E_m(0)}\rho(t)\ket{E_n(0)}$. Note that
$\inner{\Psi}{\Psi}=\Tr(\rho^2)\leq1$, i.e. this pure bipartite
state is not normalized except when the state of the open system is
pure. With these definitions, the master equation in Lindblad form
($\hbar=1$ hereafter)\cite{Lindblad}
\begin{eqnarray}
\dt{\rho}&=&-i[H,\rho]-\frac12\sum_k\left\{L_k^{\dag}L_k\rho+\rho
L_k^{\dag}L_k-2L_k\rho L_k^{\dag}\right\}\nonumber\\
&=&-i[H,\rho]+\mathcal {L}_D\rho=\mathcal
{L}\rho,\label{Lindbladeqn}
\end{eqnarray}
can be rewritten in a \schrodinger-like
equation\cite{effectivesolution}
\begin{eqnarray}
i\pt\ket{\Psi(t)}=\mathcal {H}_T\ket{\Psi(t)},\label{schrodinglike}
\end{eqnarray}
where $\mathcal {H}_T$ is the so-called effective Hamiltonian and is
defined by
\begin{eqnarray}
\mathcal{H}_T=\mathcal{H}-\mathcal{H}^A+i\sum_kL_k^AL_k,\label{effHamiltonian}
\end{eqnarray}
with $\mathcal{H}=H-\frac{i}2\sum_kL^{\dag}_kL_k$. $\mathcal {H}^A$
and $L^A_k$ are  operators for the ancilla defined by
\begin{eqnarray}
\bra{e_m(0)}O^A\ket{e_n(0)}=\bra{E_n(0)}O^{\dag}\ket{E_m(0)}.\label{operatordefinition}
\end{eqnarray}
In Eq.(\ref{Lindbladeqn}),  $H$ is the free Hamiltonian of the
system, $L_k$ are the Lindblad operators, and $\mathcal {L}$ is the
Lindblad superoperator. The eigenoperators of the Lindblad
superoperator $\mathcal{L}$ compose  the damping
basis\cite{dampingbasis}, namely a  damping basis vector satisfies
the following eigenequation,
\begin{eqnarray}
\mathcal {L}A_{\nu}=\lambda_{\nu} A_{\nu}.\label{dampingbasis}
\end{eqnarray}
Note that the eigenvalues might be complex because the Lindblad
superoperator is not Hermitian in general. One can also define the
left eigenoperator for the Lindblad superoperator,
\begin{eqnarray}
B_{\nu}\mathcal {L}=\lambda_{\nu} B_{\nu}, \label{dual}
\end{eqnarray}
which constructs  the dual damping basis. The damping basis and its
dual have the same eigenvalue  and satisfy the orthonormal condition
\begin{eqnarray}
\Tr\left(A_{\mu}B_{\nu}\right)=\delta_{\mu\nu}.
\end{eqnarray}
Now we establish the relation between the  damping basis and the
eigenstates of the effective Hamiltonian. Defining
\begin{eqnarray}
\ket{R_\nu}=\sum_{m,n=1}^NA_{mn}^{\nu}\ket{E_m(0)}\ket{e_n(0)},\label{defde}
\end{eqnarray}
where $A_{mn}^{\nu}=\langle E_m(0)|A_{\nu}|E_n(0)\rangle$. It is
easy to check that Eq.(\ref{dampingbasis}) can be rewritten as,
\begin{eqnarray}
-i\mathcal{H}_T\ket{R_{\nu}}=\lambda_{\nu}\ket{R_{\nu}}.
\end{eqnarray}
This means that the pure bipartite state $\ket{R_{\nu}}$ defined  by
the damping basis $A_{\nu}$ through Eq.(\ref{defde}) is an
eigenstate of the non-Hermitian operator $-i\mathcal {H}_T$ with
eigenvalue $\lambda_{\nu}$. For the dual damping basis $B_{\nu}$,
the relation is similar,
\begin{eqnarray}
\bra{L_{\nu}}=\sum_{m,n=0}^N\bra{e_m(0)}\bra{E_n(0)}B_{mn}^\nu,
\end{eqnarray}
and
\begin{eqnarray}
\bra{L_{\nu}}(-i\mathcal{H}_T)=\bra{L_{\nu}}\lambda_{\nu}.
\end{eqnarray}
This indicates  that the left vector $\bra{L_{\nu}}$ defined by  the
dual damping basis is the left eigenstate of the non-Hermitian
operator $-i\mathcal{H}_T$. So the orthogonal and normalized
condition for the damping basis can be represented in terms of the
left and right eigenvectors of the effective Hamiltonian,
\begin{eqnarray}
\inner{L_{\nu}}{R_{\mu}}=\delta_{\nu\mu}.
\end{eqnarray}
To shed more light on this connection, we examine  the steady state
solution of Eq.(\ref{Lindbladeqn}) in  terms of both the damping
basis and the eigenstates of the  effective Hamiltonian. The steady
state means $\dt\rho=0$, which leads to $\mathcal{L}\rho=0$, i.e.
the steady state is a  damping basis vector of the Lindblad
superoperator $\mathcal{L}$ with  zero eigenvalue. Note  we assume
$\mathcal{L}$  time-independent. In terms of the effective
Hamiltonian, this reads $\mathcal{H}_T\ket{R}=0\ket{R}$. The steady
state corresponds to the right eigenvector of the effective
Hamiltonian $\mathcal{H}_T$ with zero eigenvalue. Note the existence
of such a solution for the master equation requires  that the
determinant of the effective Hamiltonian $\mathcal{H}_T$ is zero.

With these knowledge, we now define the adiabatic and non-adiabatic
geometric phase, and formulate the dynamically decoherence-free
subspaces\cite{karasik08} by this effective Hamiltonian approach. To
define a non-adiabatic geometric phase for an open system, we define
a dynamical invariant operator $\mathcal {I}=\mathcal {I}(t)$ for
the open system by
\begin{eqnarray}
i\pt \mathcal {I}-[\mathcal{H}_T,\mathcal{I}]=0,
\end{eqnarray}
where $\mathcal{H}_T$ is the effective Hamiltonian of the open
system. This definition has the same form as the dynamical invariant
for a closed system, however the invariant operator $\mathcal {I}$
for open system  is not Hermitian in general. Assume a right(left)
basis $\{|r_i\rangle\}$ ($\{|l_i\rangle\}$) spanned by the
right(left) eigenstates of $\mathcal {I}$ exists, the wavefunction
of the composite system can be expanded in the right basis,
\begin{eqnarray}
\ket{\Psi}=\sum_jc_j(t)\ket{r_j}.\label{expansion}
\end{eqnarray}
By insetting Eq.(\ref{expansion}) into the \schrodinger-like
equation, we obtain
\begin{eqnarray}
i\dot{c_j}(t)=\sum_kc_k\bra{l_j}\mathcal{H}_T
\ket{r_k}-i\sum_kc_k\inner{l_j}{\dot{r}_k}.\label{cjdot}
\end{eqnarray}
Now we prove that for non-degenerate eigenstates $|r_j\rangle, \
(j=1,2,...)$ of $\mathcal {I}$,
\begin{eqnarray}
\bra{l_i}\mathcal{H}_T-i\pt\ket{r_j}=0,
\end{eqnarray}
for all $i\neq j$. As $|r_j\rangle$ is a right eigenstate of
$\mathcal {I}$, we have
\begin{eqnarray}
\mathcal{I}(t)\ket{r_j}=\lambda_j(t)\ket{r_j},\label{ieigen}
\end{eqnarray}
where $\lambda_j(t)$ denotes the eigenvalue corresponding to the
eigenstate $|r_j\rangle$. The derivative  of Eq.(\ref{ieigen}) with
respect to time yields($\lambda_j=\lambda_j(t)$),
\begin{eqnarray}
\bra{l_i}\dot{\mathcal{I}}\ket{r_j}=
\dot{\lambda}_j\delta_{ij}+(\lambda_j-\lambda_i)\inner{l_i}{\dot{r}_j}.\label{idot1}
\end{eqnarray}
On the other hand, from the definition of the dynamical invariant
$\mathcal {I}$, we have
\begin{eqnarray}
i\bra{l_m}\dot{\mathcal{I}}(t)\ket{r_n}=
(\lambda_n-\lambda_m)\bra{l_m}\mathcal{H}_T\ket{r_n},\label{idot2}
\end{eqnarray}
Eq.(\ref{idot1}) and Eq.(\ref{idot2}) together yield,
\begin{eqnarray}
\dot{\lambda}_j\delta_{ij}=
(\lambda_i-\lambda_j)\bra{l_i}\left(i\mathcal{H}_T+\pt\right)\ket{r_j}
\end{eqnarray}
leading to,
\begin{eqnarray}
\bra{l_i}\left(\mathcal{H}_T-i\pt\right)\ket{r_j}=0
\end{eqnarray}
for $i\neq j$ and $\lambda_i\neq \lambda_j.$ Observe that for $i\neq
j$, we get $\dot{\lambda}_j=0,\ \ \forall j$  which indicates that
the dynamical invariant has time-independent eigenvalues, similar to
the case of closed systems.  With this result, Eq.(\ref{cjdot})
reduces to,
\begin{eqnarray}
i\dot{c}_j(t)=c_j\bra{l_j}\left(\mathcal{H}_T-i\pt\right)\ket{r_j}.\label{cjdots}
\end{eqnarray}
integrating Eq.(\ref{cjdots}), we get
\begin{eqnarray}
c_j(t)=c_j(0)e^{-i\int_0^t\bra{l_j}\mathcal{H}_T\ket{r_j}dt'}
e^{-\int_0^t\bra{l_j}\frac{\partial}{\partial
t'}\ket{r_j}dt'}.\label{cjt}
\end{eqnarray}
This result tells us that when the system is initially in a right
eigenstate of $\mathcal {I}$, the system will remain to that
eigenstate up to a phase  factor at later times. The first
exponential in Eq.(\ref{cjt}) gives the dynamical phase, while the
second exponential generates the geometric phase. In a cyclic
evolution,
\begin{eqnarray}
\gamma_g^j=i\int_0^T\bra{l_j}\frac{\partial}{\partial
t'}\ket{r_j}dt'.
\end{eqnarray}
This definition was shown\cite{sarandy07} to be gauge invariant and
hence it is observable. For noncyclic evolution, a term
$\text{arg}[\langle l_i(0)|r_i(T)\rangle]$ has to be added to
$\gamma_g^i$, ensuring that geometric is phase gauge invariant. The
limitation of the present definition is that we have assumed the
Jordan blocks of $\mathcal {I}$ are one dimensional and the
eigenstates are nondegenerate. Beside, the success of our
representation relies on the existence of the dynamical invariant
$\mathcal {I}$. For discussions beyond these limitations, we refer
the reader to \cite{sarandy05,sokolov06}. Observe that the phases
defined above are for nonadiabatic evolution, since no adiabaticity
requirement has been imposed in any step of the derivation. The
adiabatic condition can exhibit in the dynamical invariant
$\mathcal{I}(t)$, for a slowly varying $\mathcal{I}(t)$ satisfying
$\frac{\partial \mathcal{I}(t)}{\partial t} \approx 0$, we obtain
$[\mathcal{H}_T,\mathcal{I}(t)]\approx 0.$ This implies that
$\mathcal{I}(t)$ and $\mathcal{H}_T$ have a common basis of
eigenstates. Thus the geometric phase defined above may be expressed
in terms of eigenstates $|R_j(t)\rangle (j=1,2,...)$ of
$\mathcal{I}(t)$ as
\begin{equation}
\gamma_g^j=i\int_0^T\langle L_j|\frac{\partial}{\partial
t}|R_j\rangle dt.
\end{equation}
This is the definition for geometric phases in an adiabatic
evolution\cite{effectiveadiabatic}.

Decoherence remains the most important obstacle to experimental
realizations of quantum processors.  One well-developed method of
counteracting the effects of decoherence is to encode quantum
information into decoherence-free subspaces
(DFSs)~\cite{pal96,duan97,duan98,zanardi97,lidar98,knill00}. DFSs
have been defined as a set of states that undergo unitary evolution
in the presence of couplings to the environment. However, unitary
evolution of a quantum state can arise in a number of ways and this
fact has resulted in different but related definitions for DFSs in
the literature. In the context of Markovian master equation, DFSs
have frequently been defined as a collection of states $\rho(t)$
fulfilling $\mathcal{L}_D \rho(t)=0$ with decoherence superoperator
$\mathcal{L}_D$  given in Eq.(\ref{Lindbladeqn}).  Most recently,
the concept of DFSs has been extended  and a dynamically stable
decoherence-free subspaces(DDFSs) has been defined\cite{karasik08}
by
\begin{eqnarray}
\pt \Tr[\rho^2(t)]=0 \  \forall t\geq 0, \ \ \mbox{with} \ \
\Tr[\rho^2(0)]=1,\label{defddfss}
\end{eqnarray}
As shown in Ref.\cite{karasik08}, Eq.(\ref{defddfss}) is a
sufficient and necessary condition for the quantum state $\rho(t)$
to evolve unitarily.  The definition Eq.(\ref{defddfss}) together
with Eq.(\ref{Lindbladeqn}) yield,
\begin{eqnarray}
\pt\Tr[\rho^2(t)]=2\langle\mathcal{L}_D\rho(t)\rangle.
\end{eqnarray}
Clearly, the condition for DDFSs is less restrictive than that for
DFSs. In the context of effective Hamiltonian approach, we have
$\pt\Tr[\rho^2(t)]=\pt\inner{\Psi(t)}{\Psi(t)}$, then a state in
DDFSs should fulfill
\begin{eqnarray}
\pt\inner{\Psi(t)}{\Psi(t)}=0, \ \forall t\geq 0, \ \ \mbox{with}\ \
\inner{\Psi(0)}{\Psi(0)}=1,
\end{eqnarray}
where $|\Psi(t)\rangle$ denotes the wavefunction of the composite
system (system+ancilla) defined by Eq.(\ref{map}). We now show that
a space spanned by
\begin{eqnarray}
H_{DDFSs}=\{\ket{\Phi_1},\ket{\Phi_2},...,\ket{\Phi_m}\}
\end{eqnarray}
is a DDFSs if and only if the basis vector fulfill,
\begin{eqnarray}
L_k\ket{\Phi_l}=\beta_k\ket{\Phi_l}\label{cond0}
\end{eqnarray}
for all $l=1,2,3,...,m$ and for all $L_k$, as well as $H_{DDFSs}$ is
invariant under
\begin{eqnarray}
\mathcal{H}_{DDFSs}&=&H-H^A -\frac
i2\sum_k\beta_kL_k^{\dag}\nonumber\\
 &+&\frac
i2\sum_k\beta_k^{\ast}L_k^{A\dagger}+i\sum_k\beta_kL_k^A.\label{freeH}
\end{eqnarray}
By using the \schrodinger-like equation, we get from
\begin{eqnarray}
\pt\inner{\Psi(t)}{\Psi(t)}=0
\end{eqnarray}
that
\begin{equation}
\langle\psi(t)| \sum_k L_k^{\dagger}L_k+ L_k^{A\dagger}L_k^A-\sum_k
(L_k^{\dagger}L_k^{A\dagger}+\text{H.c.})|\psi(t)\rangle=0.\label{cond}
\end{equation}
We now prove that $L_k\ket{\Phi_l}=\beta_k\ket{\Phi_l}$ is
equivalent to $\bra{\Phi_l}L_k^{A\dagger}=\bra{\Phi_l}\beta_k$,
where $|\Phi_l\rangle$ is a basis vector in $H_{DDFSs}.$ By
definition, any basis vector $|\Phi_l\rangle \in H_{DDFSs}$ fulfills
$L_k\ket{\Phi_l}=\beta_k\ket{\Phi_l}$, this equation can be
rewritten as
\begin{eqnarray}
\sum_{m,n}\ket{E_m(0)}\bra{E_n(0)}\bra{E_m(0)}
L_k\ket{E_n(0)}\ket{\Phi_l}=\beta_k\ket{\Phi_l}. \label{mediaeqn}
\end{eqnarray}
$|\Phi_l\rangle$ can be expanded in the basis of the composite
system as,
\begin{eqnarray}
\ket{\Phi_l}=\sum_{p,q}\Phi_l^{pq}\ket{E_p(0)}\ket{e_q(0)},\label{expanphi}
\end{eqnarray}
where $\Phi_l^{pq}$ is the expansion coefficient.  Substituting
Eq.(\ref{expanphi}) into Eq.(\ref{mediaeqn}) and noting
$\Phi_l^{pq*}=\Phi_l^{qp}$, we get $\langle
\Phi_l|L_k^{A\dagger}=\langle \Phi_l|\beta_k.$ In the last
derivation, we have used Eq.(\ref{operatordefinition}). With
$L_k\ket{\Phi_l}=\beta_k\ket{\Phi_l}$ and $\langle
\Phi_l|L_k^{A\dagger}=\langle \Phi_l|\beta_k$, we have
$\pt\Tr[\rho^2(t)]= \langle\Psi(t)| \sum_k L_k^{\dagger}L_k+
L_k^{A\dagger}L_k^A-\sum_k
(L_k^{\dagger}L_k^{A\dagger}+\text{H.c.})|\Psi(t)\rangle=0,$ where
$|\Psi(t)\rangle$ is an arbitrary state in  $\mathcal{H}_{DDFSs}$
that can be written as $|\Psi(t)\rangle=\sum_l
c_l(t)|\Phi_l\rangle.$ Thus $H_{DDFSs}$ is a dynamically
decoherence-free spaces. Note that with
$L_k\ket{\Phi_l}=\beta_k\ket{\Phi_l}$, the \schrodinger-like
equation becomes $i\frac {\partial}{\partial
t}|\Psi(t)\rangle=\mathcal{H}_{DDFSs}|\Psi(t)\rangle,$ so
$H_{DDFSs}$ being invariant under the Hamiltonian
$\mathcal{H}_{DDFSs}$ is as  required. To prove that the conditions
are necessary, we suppose that $\pt\inner{\Psi(t)}{\Psi(t)}=0$
holds. In general $L_k|\Psi(t)\rangle$ can be written as
($|\Psi(t)\rangle=|\Psi\rangle$)
\begin{eqnarray}
L_k\ket{\Psi}=\beta_k{|\Psi\rangle}+\ket{\Psi_k^{\bot}},
\end{eqnarray}
with $\ket{\Psi_k^{\bot}}$ denotes some state that is orthogonal to
the state $\ket{\Psi}$. Insetting this equation into Eq.(\ref{cond})
yields,
\begin{eqnarray}
\sum_k\inner{\Psi^{\bot}_k}{\Psi^{\bot}_k}=0.
\end{eqnarray}
Therefor any state $|\Psi(t)\rangle$ fulfilling
$\pt\inner{\Psi(t)}{\Psi(t)}=0$ certainly satisfies
Eq.(\ref{cond0}), so
 Eq.(\ref{cond0})
together with Eq.(\ref{freeH}) are the necessary condition for
DDFSs.

\section{Extension of the effective Hamiltonian approach to
non-Markovian dynamics\label{sectionnonMarkovian}}

In this section, we extend the effective Hamiltonian approach to a
non-Markovian dynamics governed by  the generalized Lindblad master
equation\cite{breuer07,nonMarkovianPRE}, the corresponding
generalization for the damping basis is also given. As mentioned
above, the generalized master equation is obtained by the projection
superoperator technique\cite{BPbookprojection,projection}. The form
(Markovian or non-Markovian) of the equation crucially depends on
the approximation used in the derivation, reflecting in the
 chosen projection superoperator. When we project the state of  the
 total system
(system plus environment) into a tensor product, we obtain the
Markovian master equation, while a non-Markovian master equation can
be obtained when we use a correlated projection. In the latter case
the master equation derived  is in the generalized Lindblad
form\cite{projection},
\begin{eqnarray}
\frac{d\rho_k}{dt}=-i[H_k,\rho_k]+\sum_{j\lambda}\left(R_{kj}^{\lambda}\rho_jR_{kj}^{\lambda\dag}-
\frac12\{R_{jk}^{\lambda\dag}R_{jk}^{\lambda}, \rho_k\}
\right),\nonumber\\ \label{Gmasterequation}
\end{eqnarray}
where $H_k$ are Hermitian operators and $R_{kj}^{\lambda}$ are
arbitrary system operators depending on the system-environment
interaction. If we only have a single component $\rho_S=\rho_1$,
this equation obviously reduces to the ordinary Markovian master
equation, whereas in general cases,  the state of the open system is
$\rho_S=\sum_k\rho_k$ with $\Tr\rho_k<1$.

To simplify  the derivation, we first rewrite
Eq.(\ref{Gmasterequation}) as
\begin{eqnarray}
i\frac{ d\rho_k}{dt}=\mathcal {H}_k\rho_k-\rho_k\mathcal
{H}_k^{\dag}+i\sum_{j\lambda}R_{kj}^{\lambda}\rho_jR_{kj}^{\lambda\dag},
\end{eqnarray}
where $\mathcal
{H}_k=H_k-\frac12i\sum_{j\lambda}R_{jk}^{\lambda\dag}R_{jk}^{\lambda}$.
Next we map this equation into a \schrodinger-like equation by
introducing an ancilla labeled by A,
\begin{widetext}
\begin{eqnarray}
i\dt{\ket{\Psi_k(t)}}=&&\sum_{m,n}\bra{E_m(0)}\mathcal {H}_k\rho_k
\ket{E_n(0)}\ket{E_m(0)}\ket{e_n(0)}-\sum_{m,n,p}\bra{E_m(0)}\rho_k\ket{E_p(0)}\bra{E_p(0)}
\mathcal{H}_k^{\dag}\ket{E_n(0)}\ket{E_m(0)}\ket{e_n(0)}\nonumber\\
&&+i\sum_{j\lambda}\sum_{m,n,p}\bra{E_m(0)}R_{kj}^{\lambda}\rho_j\ket{E_p(0)}\bra{E_p(0)}
R_{kj}^{\lambda\dag}\ket{E_n(0)}\ket{E_m(0)}\ket{e_n(0)} \nonumber\\
=&&(\mathcal{H}_k-\mathcal{H}_k^A)\sum_n\rho_k\ket{E_n(0)}\ket{e_n(0)}+
i\sum_{j\lambda}R_{kj}^{\lambda
A}R_{kj}^{\lambda}\sum_n\rho_j\ket{E_n(0)}\ket{e_n(0)}\nonumber\\
=&&(\mathcal {H}_k-\mathcal
{H}_k^A)\ket{\Psi_k(t)}+i\sum_{j\lambda}R_{kj}^{\lambda
A}R_{kj}^{\lambda}\ket{\Psi_j(t)},\label{effectivederivate}
\end{eqnarray}
\end{widetext}
where $R_{kj}^{\lambda A}$ and $\mathcal {H}_k^A$ are  operators of
the auxiliary system  defined by Eq.(\ref{operatordefinition}), and
$\ket{\Psi_k(t)}$ is the non-normalized wave function corresponding
to $\rho_k$, defined by $\ket{\Psi_k(t)}=
\sum_{m,n=1}^N\rho_{k,mn}(t)\ket{E_m(0)}\ket{e_n(0)}$ with
$\rho_{k,mn}=\langle E_m(0)|\rho_k|E_n(0)\rangle.$
Eq.(\ref{effectivederivate}) can be rewritten in a compact form,
\begin{eqnarray}
i\dt{\overrightarrow{\ket{\mathbf{\Psi}(t)}}}=
\mathbf{H}\overrightarrow{\ket{\mathbf{\Psi}(t)}},\label{scheqM}
\end{eqnarray}
where $\overrightarrow{\ket{\mathbf{\Psi}(t)}}=
\left[\ket{\Psi_1(t)},\ket{\Psi_2(t)},\cdots\right]^\mathbf{T}$ is a
wave function vector, and $\mathbf{H}$ is an effective Hamiltonian
that we will refer to  as   effective Hamiltonian matrix hereafter.
The matrix elements of this effective Hamiltonian are,
\begin{eqnarray}
\mathbf{H}_{kj}=\delta_{kj}(\mathcal {H}_k-\mathcal
{H}_k^A)+i\sum_{\lambda}R_{kj}^{\lambda
A}R_{kj}^{\lambda}.\label{effHnon}
\end{eqnarray}
Note that the elements of the effective Hamiltonian matrix are
operators and in general not non-Hermitian. The diagonal element
includes a term $\mathcal {H}_k-\mathcal {H}_k^A$, which describes
the free evolution for the open system and the ancilla, and an
interaction term $i\sum_{\lambda}R_{kk}^{\lambda
A}R_{kk}^{\lambda}$, which results in the quantum jump on
$\ket{\Psi_k}$. The off-diagonal elements represent  the coupling
between  $\ket{\Psi_k}$ $(k=1,2,3,...)$, leading the quantum jump
from one component (say $\ket{\Psi_i}$) to the other component (say
$\ket{\Psi_j}$)\cite{Huang2008PRE}. When we have only one component,
the results reduce to Ref.\cite{effectivesolution}. Assuming the
Hamiltonian of the original system is time independent, we can
obtain formally the evolution of the wave function vector
$\overrightarrow{\ket{\mathbf{\Psi}(t)}}$ as
\begin{eqnarray}
{\overrightarrow{\ket{\mathbf{\Psi}(t)}}}=
e^{-i\mathbf{H}t}{\overrightarrow{\ket{\mathbf{\Psi}(0)}}}.
\end{eqnarray}
By mapping this solution back to the density matrix through
$\ket{\Psi_k(t)}=
\sum_{m,n=1}^N\rho_{k,mn}(t)\ket{E_m(0)}\ket{e_n(0)}$ with
$\rho_{k,mn}=\langle E_m(0)|\rho_k|E_n(0)\rangle,$ we obtain the
time evolution governed by Eq.(\ref{Gmasterequation}).

Similarly,  we can define a damping basis for the generalized
Lindblad equation as a set of operators
$\{A_1^{\nu},A_2^{\nu},\cdots\}$ satisfying
$-i[H_k,A_k^{\nu}]+\sum_{j\lambda}\left(R_{kj}^{\lambda}A_j^{\nu}R_{kj}^{\lambda\dag}-
\frac12\{R_{jk}^{\lambda\dag}R_{jk}^{\lambda}, A_k^{\nu}\}
\right)=\lambda_{\nu}A_k^{\nu}$.  The definition for the dual of
this damping basis is similar. The connection of this damping basis
to the eigenstates of the effective Hamiltonian matrix will be given
through an example in the Appendix.

\section{Example\label{sectionapplication}}

In this section, we will use the model and the master equation given
in Ref.\cite{nonMarkovianPRE} to illustrate our method. Consider a
two-state system coupled to an environment. The environment consists
of a large number of energy levels which constitute  two energy
bands. The lower energy band contains $N_1$ levels while the upper
one $N_2$ levels.  A detailed description for this model can be
found in \cite{model2,model3}. In the interaction picture, the
non-Markovian master equation reads,
\begin{eqnarray}
\frac d{dt}\rho_S^{(1)}(t)=
\gamma_1\sigma^+\rho_S^{(2)}(t)
\sigma^--\frac{\gamma_2}2\{\sigma^+\sigma^-,\rho_S^{(1)}(t)\},\nonumber \\
\frac d{dt}\rho_S^{(2)}(t)=
\gamma_2\sigma^-\rho_S^{(1)}(t)
\sigma^+-\frac{\gamma_1}2\{\sigma^-\sigma^+,\rho_S^{(2)}(t)\}.\label{exaplemaster}
\end{eqnarray}
where $\gamma_i, (i=1,2)$ depend on the system-environment couplings
as well as the energy gap of the environment. $\sigma^{\pm}$ are the
Pauli operators. Defining $\Pi_1=\sum_{n_1}|n_1\rangle\langle n_1|$
and $\Pi_2=\sum_{n_2}|n_2\rangle\langle n_2|$,
$\Pi_1+\Pi_2=\mathbb{I}_E$, where the index $n_1$ labels  the level
of lower energy band, and $n_2$ is the level index  of the upper
band. The two unnormalized density matrixes can be obtained by
$\rho^{(i)}_S=\Tr_E(\Pi_i\rho_T),i=1,2$, where $\rho_T$ is the
density matrix for  the whole system (the system plus the
environment). The reduced density matrix for the system is then
given by $\rho=\rho^{(1)}_S+\rho^{(2)}_S$. Equation
(\ref{exaplemaster}) can be written in the form of
Eq.(\ref{Gmasterequation}) by setting $H_i=0, \ (i=1,2)$,
$R_{11}=R_{22}=0$, $R_{12}=\sqrt{\gamma_1}\sigmap$, and
$R_{21}=\sqrt{\gamma_2}\sigmam$. By the effective Hamiltonian
approach, this equation can be easily solved and an analytical
expression for the density matrix can be given as follows. The
elements of the effective Hamiltonian matrix for
Eq.(\ref{exaplemaster}) are,
\begin{eqnarray}
&&\mathbf{H}_{11}=-\frac12i\gamma_2(\sigmap\sigmam+\tau^+\tau^-), \nonumber\\
&&\mathbf{H}_{12}=i\gamma_1\sigmap\tau^+ ,\nonumber\\
&&\mathbf{H}_{21}=i\gamma_2\sigmam\tau^- , \nonumber\\
&&\mathbf{H}_{22}=-\frac12i\gamma_1(\sigmam\sigmap+\tau^-\tau^+),\label{ehe}
\end{eqnarray}
where $\tau^+$ and $\tau^-$ are the Pauli operators  for the
auxiliary system. The elements of the time evolution operator
$\mathbf{U}$ corresponding to this effective Hamiltonian can be
obtained by  simple algebras,
\begin{widetext}
\begin{eqnarray}
&&\mathbf{U}_{11}=\frac{\gamma_1+\gamma_2e^{-(\gamma_1+\gamma_2)t}}{\gamma_1+\gamma_2}\sigmap\sigmam\tau^+\tau^-
+e^{-\frac12\gamma_2t}(\sigmap\sigmam\tau^-\tau^++\sigmam\sigmap\tau^+\tau^-)+\sigmam\sigmap\tau^-\tau^+,\nonumber\\
&&\mathbf{U}_{12}=\frac{\gamma_1(1-e^{-(\gamma_1+\gamma_2)t})}{\gamma_1+\gamma_2}\sigmap\tau^+,\nonumber\\
&&\mathbf{U}_{21}=\frac{\gamma_2(1-e^{-(\gamma_1+\gamma_2)t})}{\gamma_1+\gamma_2}\sigmam\tau^-,\nonumber\\
&&\mathbf{U}_{22}=\sigmap\sigmam\tau^+\tau^-+e^{-\frac12\gamma_1t}(\sigmap\sigmam\tau^-\tau^++\sigmam\sigmap\tau^+\tau^-)
+\frac{\gamma_2+\gamma_1e^{-(\gamma_1+\gamma_2)t}}{\gamma_1+\gamma_2}\sigmam\sigmap\tau^-\tau^+.
\end{eqnarray}
With this evolution operator, we can solve the \schrodinger-like
equation and map the solution back to the density matrix, then
obtain
\begin{eqnarray}
\rho_S^{(1)}&&=\left(\begin{array}{cc}
\frac{(\gamma_1+\gamma_2e^{-(\gamma_1+\gamma_2)t})\rho^{(1)}_{11}(0)+\gamma_1(1-e^{-(\gamma_1+\gamma_2)t})\rho^{(2)}_{22}(0)}{\gamma_1+\gamma_2},& e^{-\frac12\gamma_2t}\rho^{(1)}_{12}(0)\\
e^{-\frac12\gamma_2t}\rho^{(1)}_{21}(0),
&\rho^{(1)}_{22}(0)\end{array} \right),\nonumber\\
\rho_S^{(2)}&&=\left(\begin{array}{cc} \rho^{(2)}_{11}(0),&e^{-\frac12\gamma_1t}\rho^{(2)}_{12}(0)\\
e^{-\frac12\gamma_1t}\rho^{(2)}_{21}(0),&
\frac{\gamma_2(1-e^{-(\gamma_1+\gamma_2)t})\rho^{(1)}_{11}(0)+(\gamma_2+\gamma_1e^{-(\gamma_1+\gamma_2)t})\rho^{(2)}_{22}(0)}{\gamma_1+\gamma_2}
\end{array}\right).
\end{eqnarray}
\end{widetext}
For an  initial state  that only the lower band is populated, this
solution reduces to the well-known result given in
Ref.\cite{nonMarkovianPRE}. We now turn to the steady state solution
of the master equation. The existence of steady states requires that
the determinant of the effective Hamiltonian matrix is zero. This
requirement is met for $\mathbf{H}$ in Eq.(\ref{ehe}). As shown in
Sec.{\rm II}, the steady states are given by  the eigenstates of the
effective Hamiltonian matrix with zero eigenvalue. It is  threefold
degenerate in this example, and the three degenerate eigenstates in
terms of  the damping basis are
\begin{eqnarray}
&&A^{01}=\left\{\rho_S^{(1)}=0,\rho_S^{(2)}=\ext{e}{e}\right\},\nonumber\\
&&A^{02}=\left\{\rho_S^{(1)}=\ext{g}{g},\rho_S^{(2)}=0\right\},\\
&&A^{03}=\left\{\rho_S^{(1)}=
\frac{\gamma_1}{\gamma_1+\gamma_2}\ext{e}{e},
\rho_S^{(2)}=\frac{\gamma_2}{\gamma_1+\gamma_2}\ext{g}{g}\right\},\nonumber
\end{eqnarray}
where $\ket{e}$ and $\ket{g}$ are the excited state and the ground
state for the two-state system, respectively. This steady state
solution tells us that there are three types of equilibrium for the
whole system.  $A^{01}$ describes an equilibrium  that the two-state
system is in its excited state while the environment occupies  the
upper band; $A^{02}$ represents an equilibrium in which the
two-state system is in its ground state with  the environment  in
its lower band; steady state $A^{03}$ indicates that the ratio of
the population in the ground state to the population  in the excited
state for the system is exactly the same as the ratio between the
populations of the environment in its lower and upper band. One can
see from these steady states that this non-Markovian master equation
indeed accounts for  the memory effect of the environment, leading
to an environment-state-dependent evolution of the system.

\section{applications: adiabatic evolution and dynamically stable
decoherence-free subspaces} In this section, we will apply the
effective Hamiltonian approach to define an adiabatic evolution for
the non-Markovian open system, and present an analysis for the
dynamically stable decoherence-free subspaces. The adiabatic
theorem\cite{adiabaticclosed} is one of the oldest and most useful
tools in quantum mechanics. It tells us that if a state is an
instantaneous eigenstate of a sufficiently slowly varying
Hamiltonian $H(t)$ at one time, then it will remain close to that
eigenstate up to a phase factor at later times, while its eigenvalue
evolves continuously. The notion of adiabaticity  has been extended
to Markovian open systems in\cite{sarandy05,effectiveadiabatic},
however no extension for the adiabaticity from Markovian dynamics to
non-Markovian dynamics can be found in the  literature.

Having the effective Hamiltonian Eq.(\ref{effHnon}), such an
extension is straightforward and the adiabaticity can be formulated
as the following.  {\it An open system govern by the master equation
Eq.(\ref{Gmasterequation}) is said to undergo adiabatic evolution if
the composite system govern by the effective Hamiltonian
$\mathbf{H}$ Eq.(\ref{effHnon}) (it may depends on time directly or
indirectly) evolves adiabatically}. Following the procedure
in\cite{effectiveadiabatic}, we arrive at the adiabatic
condition\cite{noteforadia},
\begin{eqnarray}
\left|\frac{\inner{\mathbf{L}_m(t)}{\dot{\mathbf{R}}_n(t)}}
{\lambda_m-\lambda_n}\right|\ll1, \label{adiacon}
\end{eqnarray}
where $\bra{\mathbf{L}_m(t)}$ and $\ket{\mathbf{R}_n(t)}$ are the
left and right eigenstate of the effective Hamiltonian matrix,
respectively. We would like to address that the present discussion
is restricted  to systems where the effective Hamiltonian matrix is
diagonalizable with nondegenerate eigenvalues. A generalization to
the non-diagonalizable and degenerate case can be made  by using the
methods in Ref.\cite{sokolov06}.

\begin{figure}
\includegraphics*[width=0.8\columnwidth,
height=0.6\columnwidth]{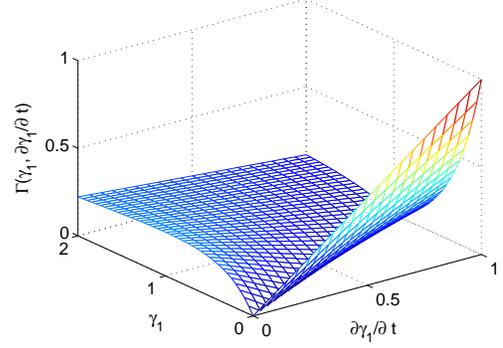} \caption{Character
function $\Gamma$ versus  $\gamma_1(T)$ and $\pt{\gamma_1}(T)$ at a
fixed time point $T$. $\gamma_1$ describes the decay rate of the
system, whereas $\pt{\gamma_1}$ characterizes the change rate of
$\gamma_1$. Both $\gamma_1$ and  $\pt{\gamma_1}$ can vary
independently, for example when $\gamma_1=\omega t$,
$\pt{\gamma_1}=\omega$. The other parameters chosen are $\gamma_2=1$
and $\pt{\gamma_2}=1.$ } \label{characterfunction}
\end{figure}

\begin{figure}
\includegraphics*[width=0.8\columnwidth,
height=0.6\columnwidth]{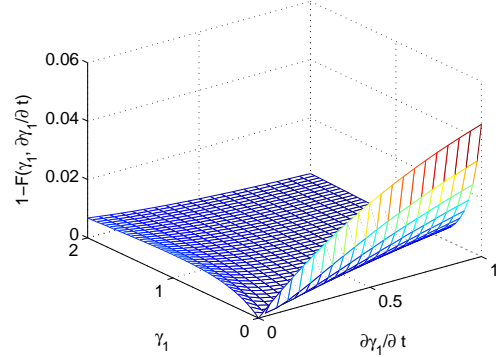} \caption{The fidelity
($1-\mathcal{F}$) as functions of $\gamma_1$ and $\pt\gamma_1.$ The
other parameters are the same as in Fig. \ref{characterfunction}}
\label{fidelity}
\end{figure}

To check whether the adiabatic condition Eq.(\ref{adiacon}) can
guarantee the adiabatic evolution in non-Markovian open systems,  we
simulate the dynamics governed by Eq.(\ref{exaplemaster}) and
compare it to the result from the adiabatic evolution defined above.
To this end, we assume that the two parameters
$\gamma_1=\gamma_1(t)$ and $\gamma_2=\gamma_2(t)$ are
time-dependent.  In fact, the time-independent parameters $\gamma_1$
and $\gamma_2$ can be obtained under special
conditions\cite{BPbookprojection}. We define the following function
\begin{eqnarray}
\Gamma=\max\left\{\left|\frac{\inner{\mathbf{L}_m(t)}{\dot{\mathbf{R}}_n(t)}}{\lambda_m-\lambda_n}\right|\right\}
\end{eqnarray}
to characterize  the violation of the adiabatic condition, where
$\max$ is taken over all $m$ and $n$, i.e. all the left and right
eigenstates of the effective Hamiltonian matrix. The detailed
expressions for the left and right eigenstates are given in the
Appendix. We plot $\Gamma$ as functions of $\gamma_1(T)$ and
$\pt{\gamma_1}(T)$ in Fig.\ref{characterfunction}. To compare the
numerical solution $\rho$ with the adiabatic evolution $\rho_a$, we
use the fidelity\cite{fidelity} as a measure to quantify the
difference between the two density matrices. The fidelity is defined
as
$\mathcal{F}(\rho,\rho_a)=\Tr\sqrt{\sqrt{\rho}\rho_a\sqrt{\rho}}$.
It reaches one when the two states are the same. In
Fig.\ref{fidelity}, we plot the fidelity ($1-\mathcal{F}$) as
functions of $\gamma_1(T)$ and $\pt\gamma_1(T)$. We choose $A^{03}$
as the initial state. The other parameters chosen in these figure
are $\gamma_2(T)=1$ and $\pt\gamma_2(T)=1$. Comparing
Fig.\ref{fidelity} with Fig.\ref{characterfunction}, we can find
that the character function $\Gamma$  and the fidelity have a very
similar appearance, i.e. the smaller the character function is, the
slighter the difference between the numerical simulation  $\rho$ and
the adiabatic result  $\rho_a$ is. This verifies the adiabatic
condition.

In the context of effective Hamiltonian approach, we now formulate
the dynamically stable decoherence-free subspaces for the
non-Markovian dynamics governed by Eq.(\ref{Gmasterequation}). In
terms of the wave function vector
$\overrightarrow{\ket{\mathbf{\Psi}(t)}}=
\left[\ket{\Psi_1(t)},\ket{\Psi_2(t)},\cdots\right]^\mathbf{T},$ the
definition of DDFSs given in Eq.(\ref{defddfss}) can be expressed as
\begin{eqnarray}
&&\sum_{i,j}\pt\inner{\Psi_i(t)}{\Psi_j(t)}=0, \ \forall t\geq 0, \
\
\nonumber\\
&& \mbox{with}\ \ \sum_{i,j}\inner{\Psi_i(0)}{\Psi_j(0)}=1.
\end{eqnarray}
By the \schrodinger-like equation Eq.(\ref{scheqM}), this definition
immediately follows,
\begin{equation}
\sum_{ijk}\langle \Psi_i|
\mathbf{H}_{jk}-\tilde{{\mathbf{H}}}^*_{ji}|\Psi_k\rangle=0.
\end{equation}
If
$R_{kj}^{\lambda}|\Phi_m\rangle=\beta_{k}^{\lambda}|\Phi_m\rangle$
holds for all $\lambda, k,j$ and the eigenvalue $\beta_k^{\lambda}$
is independent of $m$ and $j$, we have
$\sum_{i,j}\pt\inner{\Psi_i(t)}{\Psi_j(t)}=0,$ namely
$\mathcal{H}^{\prime}_{DDFSs}\equiv \{|\Phi_m\rangle \}$ composes a
DDFSs. With
$R_{kj}^{\lambda}|\Phi_m\rangle=\beta_{k}^{\lambda}|\Phi_m\rangle$,
the elements of the effective Hamiltonian matrix reduce to
\begin{eqnarray}
H_{kj}^{DDFSs}&=&\delta_{kj}(H_k-\frac i
2\sum_{p\lambda}\beta_{p}^{\lambda} R_{pk}^{\lambda
\dagger}-H_k^{A}+\frac i
2\sum_{p\lambda}\beta_{p}^{\lambda *}R_{pk}^{\lambda A \dagger})\nonumber\\
&+&i\sum_{\lambda}\beta_{k}^{\lambda}R_{kj}^{ \lambda A}.
\label{reduH}
\end{eqnarray}
Thus that $\mathcal{H}^{\prime}_{DDFSs}$ is invariant under the
effective Hamiltonian with matrix  elements $H_{kj}^{DDFSs}$ is
required. We observed that these conditions are similar to that in
Eqs (\ref{cond0},\ref{freeH}), so by repeating the same procedure,
we can prove that these conditions are necessary for DDFSs.

\section{Conclusion and Discussion\label{sectionconclusion}}
Based on the effective Hamiltonian approach, we have presented  a
self-consistent framework for the analysis of geometric phases and
dynamically stable decoherence-free subspaces in open systems.
Comparisons to the earlier works are made. A connection of this
effective Hamiltonian approach to the method based on the damping
basis has been established.  This effective Hamiltonian approach is
then extended to a non-Markovian case with the generalized Lindblad
master equation. As an example, the non-Markovian  master equation
describing a dissipative two-level system has been solved by this
method. An adiabatic evolution has been  defined and the
corresponding adiabatic condition has been given based on this
extended effective Hamiltonian approach. A necessary and sufficient
condition for the dynamically stable decoherence-free space is also
presented. The effective Hamiltonian approach can be extended to all
master equations which are local in time.  The geometric phase
defined through the effective Hamiltonian is in fact a difference of
two geometric phases, when the system under consideration is a
closed system with pure initial states. The present analysis is
available for all master equations that are local in time, i.e., at
any point in time the future evolution only depends on the present
state and not on the history of the system. We restrict ourself in
this paper to consider the case where the effective Hamiltonian is
diagonalizable and its eigenstates are nondegenerate. The situation
beyond this limitation can be analyzed by introducing Jordan blocks
in the Hamiltonian, which beyond the scope of this paper.

This work is supported NUS Research Grant WBS:R-710-000-008-271 and
NSF of China under grant Nos. 60578014 and 10775023.\\

\section*{Appendix: Eigenstates and The Corresponding Eigenvalues of The Effective
Hamiltonian Matrix} In this section, we list the eigenstates and the
corresponding eigenvalues
 for the effective Hamiltonian matrix $\mathbf{H}$. The
eight eigenvalues are, $0$ (threefold degenerate, corresponding to
$A^{01},A^{02},A^{03}$), $-\frac{\gamma_1}2i$ (twofold degenerate
corresponding to $A^{11},A^{12}$), $-\frac{\gamma_2}2i$ (twofold
degenerate corresponding to $A^{21},A^{22}$), and
$-(\gamma_1+\gamma_2)$ (corresponding to $A^{3}$). The corresponding
right eigenstates are (in damping basis)
\begin{eqnarray}
&&A^{01}=\left\{A^{01}_1=0,A^{01}_2=\ext{e}{e}\right\},\nonumber\\
&&A^{02}=\left\{A^{02}_1=\ext{g}{g},A^{02}_2=0\right\},\nonumber\\
&&A^{03}=\left\{A^{03}_1=\frac{\gamma_1}{\gamma_1+\gamma_2}\ext{e}{e},A^{03}_2=\frac{\gamma_2}{\gamma_1+\gamma_2}\ext{g}{g}\right\},\nonumber\\
&&A^{11}=\left\{A^{11}_1=0,A^{11}_2=\ext{g}{e}\right\}, \nonumber\\
&&A^{12}=\left\{A^{12}_1=0,A^{12}_2=\ext{e}{g}\right\},\nonumber\\
&&A^{21}=\left\{A^{21}_1=\ext{g}{e},A^{21}_2=0\right\},\nonumber\\
&&A^{22}=\left\{A^{22}_1=\ext{e}{g},A^{22}_2=0\right\},\nonumber\\
&&A^{3}=\left\{A^{3}_1=\ext{e}{e},A^{3}_2=-\ext{g}{g}\right\},\nonumber
\end{eqnarray}
and their dual
\begin{eqnarray}
&&B^{01}=\left\{B^{01}_1=0,B^{01}_2=\ext{e}{e}\right\}, \nonumber\\
&&B^{02}=\left\{B^{02}_1=\ext{g}{g},B^{01}_2=0\right\}, \nonumber\\
&&B^{03}=\left\{B^{03}_1=\ext{e}{e},B^{03}_2=\ext{g}{g}\right\}, \nonumber\\
&&B^{11}=\left\{B^{11}_1=0,B^{01}_2=\ext{e}{g}\right\}, \nonumber\\
&&B^{12}=\left\{B^{12}_1=0,B^{12}_2=\ext{g}{e}\right\}, \nonumber\\
&&B^{21}=\left\{B^{21}_1=\ext{e}{g},B^{21}_2=0\right\}, \nonumber\\
&&B^{22}=\left\{B^{22}_1=\ext{g}{e},B^{22}_2=0\right\}, \nonumber\\
&&B^{3}=\left\{B^{3}_1=\frac{\gamma_2}{\gamma_1+\gamma_2}\ext{e}{e},B^{3}_2=-\frac{\gamma_1}{\gamma_1+\gamma_2}\ext{g}{g}\right\}.
\nonumber
\end{eqnarray}
The alternative expressions can be given in the  Hilbert space
spanned by $\{|ee\rangle,|eg\rangle,|ge\rangle,|gg\rangle\}$
\begin{eqnarray}
&&A^{01}_1=\left[0,0,0,0\right]^{\mathbf{T}},\  A^{01}_2=\left[1,0,0,0\right]^{\mathbf{T}}, \nonumber\\
&&A^{02}_1=\left[0,0,0,1\right]^{\mathbf{T}},\ A^{02}_2=\left[0,0,0,0\right]^{\mathbf{T}}, \nonumber\\
&&A^{03}_1=\left[\frac{\gamma_1}{\gamma_1+\gamma_2},0,0,0\right]^{\mathbf{T}},\
 A^{03}_2=\left[0,0,0,
\frac{\gamma_2}{\gamma_1+\gamma_2}\right]^{\mathbf{T}},\nonumber\\
&&A^{11}_1=\left[0,0,0,0\right]^{\mathbf{T}},\ A^{11}_2=\left[0,0,1,0\right]^{\mathbf{T}}, \nonumber\\
&&A^{12}_1=\left[0,0,0,0\right]^{\mathbf{T}},\ A^{12}_2=\left[0,1,0,0\right]^{\mathbf{T}}, \nonumber\\
&&A^{21}_1=\left[0,0,1,0\right]^{\mathbf{T}},\ A^{21}_2=\left[0,0,0,0\right]^{\mathbf{T}}, \nonumber\\
&&A^{22}_1=\left[0,1,0,0\right]^{\mathbf{T}},\ A^{22}_2=\left[ 0,0,0,0\right]^{\mathbf{T}}, \nonumber\\
&&A^{3}_1=\left[1,0,0,0\right]^{\mathbf{T}},\
A^{3}_2=\left[0,0,0,-1\right]^{\mathbf{T}}, \nonumber
\end{eqnarray}
are the right eigenstates, and
\begin{eqnarray}
&&B^{01}_1=\left[0,0,0,0\right],\ B^{01}_2=\left[1,0,0,0\right], \nonumber\\
&&B^{02}_1=\left[0,0,0,1\right],\ B^{02}_2=\left[0,0,0,0\right],\nonumber\\
&&B^{03}_1=\left[1,0,0,0\right],\ B^{03}_2=\left[0,0,0,1\right], \nonumber\\
&&B^{11}_1=\left[0,0,0,0\right],\ B^{11}_2=\left[0,0,1,0\right], \nonumber\\
&&B^{12}_1=\left[0,0,0,0\right],\ B^{12}_2=\left[0,1,0,0\right], \nonumber\\
&&B^{21}_1=\left[0,0,1,0\right],\ B^{21}_2=\left[0,0,0,0\right], \nonumber\\
&&B^{22}_1=\left[0,1,0,0\right],\ B^{22}_2=\left[0,0,0,0\right], \nonumber\\
&&B^{3}_1=\left[\frac{\gamma_2}{\gamma_1+\gamma_2},0,0,0\right],
B^{3}_2=\left[0,0,0,-\frac{\gamma_1}{\gamma_1+\gamma_2}\right],
\nonumber
\end{eqnarray}
are the left eigenstates.
\end{document}